\documentclass[11pt,thmsb,fleqn]{article}%
\usepackage{graphicx}
\usepackage{amsmath}%
\usepackage{amsfonts}%
\usepackage{amssymb}
\newtheorem{theorem}{Theorem}[section]

\newtheorem{corollary}[theorem]{Corollary}

\newtheorem{definition}[theorem]{Definition}

\newtheorem{proposition}[theorem]{Proposition}

\newenvironment{proof}[1][Proof]{\textbf{#1.} }{\ \rule{0.5em}{0.5em}}
\begin{document}

\title{The Lattice of Fuzzy Intervals \\and Sufficient Conditions for Its Distributivity }
\author{Ath. Kehagias}
\maketitle

\begin{abstract}
Given a \emph{reference} lattice $(X,\sqsubseteq)$, we define \emph{fuzzy
intervals }to be the fuzzy sets such that their $p$-cuts are crisp closed
intervals of $(X,\sqsubseteq)$. We show that: given a complete lattice
$(X,\sqsubseteq)$ the collection of its fuzzy intervals is a complete lattice.
Furthermore we show that: if $(X,\sqsubseteq)$ is completely distributive then
the lattice of its fuzzy intervals is distributive.

\vspace{0.5cm}


\end{abstract}

\section{Introduction}

\label{sec01}

The following is a small sample of the large literature on fuzzy algebras.
Rosenfeld wrote the first paper on \emph{fuzzy groups }\cite{Ros01}; a recent
review is \cite{AjGrp01}. \emph{Fuzzy rings }and \emph{fuzzy ideals of rings
}are studied in \cite{Yue01,Dixit01,AjRng01,ZahRng01}. Seselja, Tepavcevska
and others have presented a far reaching famework of L-fuzzy and P-fuzzy
algebras \cite{Ses01,Ses03,Ses04}. 

\emph{Fuzzy lattices} are a particular type of fuzzy algebras. A fuzzy lattice
is a fuzzy set such that its cuts are sublattices of a ``reference lattice''
$(X,\sqsubseteq)$. Relatively little has been published on fuzzy lattices.
Yuan and Wu introduced the concept \cite{Yuan1} and Ajmal studied it in
greater detail \cite{AjLat01}. Swamy and Raju \cite{Swamy1} and, more
recently, Tepavcevska and Trajkovski \cite{Tep01} studied \emph{L-fuzzy
lattices}.\footnote{Two additional senses of the term ``fuzzy lattice'' should
also be mentioned. Kaburlasos and Petridis use \emph{fuzzy inclusion measures}
\cite{VGK03,VGK01,VGK02} to introduce a concept of ``fuzzy lattice'' which is
different from the one used in the previously mentioned works; however there
is an interesting connection between the two approaches through the concept of
\emph{fuzzy orders}. In addition, \cite{Wang01,Wang03,Wang02} and many others
use the term ``fuzzy lattice'' to denote a quite different mathematical
concept, namely a completely distributive lattice with an order reversing
involution.}.

In this note we introduce \emph{fuzzy intervals }within the context of fuzzy
lattices. I.e. a fuzzy interval is defined to be a fuzzy set such that its
cuts are closed intervals of a reference lattice $(X,\sqsubseteq)$. It appears
that fuzzy intervals (in this lattice theoretic sense) have not been studied
previously. A special case which has been extensively studied is that of fuzzy
intervals with the reference lattice $(X,\sqsubseteq)$ being a set of
\emph{real numbers }\cite{Nguyen}. Some connections between this special case
and the more general case studied here will be discussed briefly in Section
\ref{sec05}..

As mentioned, our study of fuzzy intervals is lattice theoretic. We establish
some basic properties of fuzzy intervals and we show the following: given a
complete lattice $(X,\sqsubseteq)$, the collection of its fuzzy intervals is a
complete lattice; if $(X,\sqsubseteq)$ is \emph{completely distributive} then
the lattice of its fuzzy intervals is distributive.

\section{Preliminaries}

\label{sec02}

In what follows, the closed unit interval is denoted by $L\doteq
\lbrack0,1]\subseteq R$. The usual order of real numbers is denoted by $\leq$;
the maximum (resp. minimum) of $x,y$ is denoted by $x\vee y\ $\ (resp.
$x\wedge y $). Given a set $P\subseteq L$, $\vee P$ (resp. $\wedge P$) denotes
the supremum (resp. the infimum)\ of $P$. $(L,\leq,\vee,\wedge)$ is a totally
ordered set.

The \emph{reference lattice }is denoted by $(X,\sqsubseteq,\sqcup,\sqcap)$ and
it is assumed to be complete. Hence, for every $Y\subseteq X$ the elements
$\sqcap Y$, $\sqcup Y$ exist; in particular, there exist $\sqcap X$ (the
minimum element of $X$) and $\sqcup X$ (the maximum element of $X$), hence we
can write $X=[\sqcap X,\sqcup X]$.

\begin{definition}
\label{cnt0201}A \emph{fuzzy set} is a function $M:X\rightarrow L$. The
collection of all fuzzy sets (from $X$ to $L$) will be denoted by
$\mathbf{F}(X,L)$ or simply by $\mathbf{F}.$
\end{definition}

In a standard manner, we introduce an order on $\mathbf{F}$ using the
``pointwise'' order of $(L,\leq,\vee,\wedge)$. The symbols $\leq,\vee,\wedge$
will be used without danger of confusion.

\begin{definition}
\label{cnt0202}For $M,N\in\mathbf{F}$ we write $M\leq N$ iff for all $x\in X$
we have: $M(x)\leq N(x)$.
\end{definition}

\begin{definition}
\label{cnt0203}For $M,N\in\mathbf{F}$: we define the fuzzy set $M\vee N$ by:
$(M\vee N)(x)\doteq M(x)\vee N(x)$; we define the fuzzy set $M\wedge N$ by:
$(M\wedge N)(x)\doteq M(x)\wedge N(x)$.
\end{definition}

It is well known \cite{Nguyen} that $\leq$ is an order on $\mathbf{F}$ and
that $(\mathbf{F},\leq,\vee,\wedge)\ $is a complete and distributive lattice
with $\sup(M,N)=M\vee N$, $\inf(M,N)=M\wedge N$.

\begin{definition}
\label{cnt0204}Given a fuzzy set $M:X\rightarrow L$, the $p$-\emph{cut }of $M
$ is denoted by $M_{p}$ and defined by $M_{p}\doteq\{x:M(x)\geq p\}.$
\end{definition}

We will need some properties of $p$-cuts, summarized in the following
propositions. Their proofs can be found in \cite{Nguyen}.

\begin{proposition}
\label{cnt0205}Take any $M\in\mathbf{F}$ with $p$-cuts $\left\{
M_{p}\right\}  _{p\in L}$ and $N\in\mathbf{F}$ with $p$-cuts $\left\{
N_{p}\right\}  _{p\in L}$. Then $M=N$ iff $\ $for all $p\in L$ we have
$M_{p}=N_{p}$.
\end{proposition}

\begin{proposition}
\label{cnt0206}Take any $M\in\mathbf{F}$ with $p$-cuts $\left\{
M_{p}\right\}  _{p\in L}$. Then we have the following.

(i)\ For all $p,q\in L$ we have: $p\leq q\Rightarrow M_{q}\subseteq M_{p}$.

(ii)\ For all $P\subseteq L$ we have: $\cap_{p\in P}M_{p}=M_{\vee P}$.

(iii) $M_{0}=X$.
\end{proposition}

\begin{proposition}
\label{cnt0207}Consider a family of sets $\{\widetilde{M}_{p}\}_{p\in L}$
which satisfy the following.

(i)$\;$For all $p,q\in L$ we have: $p\leq q\Rightarrow\widetilde{M}%
_{q}\subseteq\widetilde{M}_{p}$.

(ii)\ For all $P\subseteq L$ we have: $\cap_{p\in P}\widetilde{M}%
_{p}=\widetilde{M}_{\vee P}$.

(iii) $\widetilde{M}_{0}=X$.

Define the fuzzy set $M(x)=\vee\{p:x\in\widetilde{M}_{p}\}$. Then for all
$p\in L$ we have $M_{p}=\widetilde{M}_{p}.$
\end{proposition}

We will also need some well-known properties of (crisp)\ closed intervals in a lattice.

\begin{definition}
\label{cnt0208}Given $x_{1},x_{2}\in X$, with $x_{1}\sqsubseteq x_{2}$, the
\emph{closed interval} $[x_{1},x_{2}]$ is defined by $[x_{1},x_{2}]\doteq$
$\{z:x_{1}\sqsubseteq z\sqsubseteq x_{2}\}.$
\end{definition}

We consider the empty set $\emptyset$ to be a closed interval, the so called
\emph{empty} \emph{interval}. This can also be denoted as $[x_{1},x_{2}]$ with
any $x_{1},x_{2}$ such that $x_{1}\not \sqsubseteq x_{2}$. Denote by
$\mathbf{I}\ $\ the collection of (crisp) closed intervals of $X$ (including
the empty interval). The structure $(\mathbf{I},\subseteq)$ is an ordered set.
In fact it is a lattice, as the following propositions show (proofs are
omitted for brevity; they follow from the fact that being a closed interval is
a \emph{closure property} on $(\mathbf{I},\subseteq)$ \cite{Birkhoff}).

\begin{proposition}
\label{cnt0209}Given any nonempty interval $A=[a_{1},a_{2}]$ $\subseteq X$, we
have $a_{1}=\sqcap A$, $a_{2}=\sqcup A$.
\end{proposition}

\begin{proposition}
\label{cnt0210}Given any family of closed intervals $\mathbf{J\subseteq I}$
the set $\cap_{\lbrack a_{1},a_{2}]\in\mathbf{J}}[a_{1},a_{2}]$ is a closed
interval; more specifically, we have
\[
\cap_{\lbrack a_{1},a_{2}]\in\mathbf{J}}[a_{1},a_{2}]=[\sqcup_{\lbrack
a_{1},a_{2}]\in\mathbf{J}}a_{1},\sqcap_{\lbrack a_{1},a_{2}]\in\mathbf{J}%
}a_{2}]
\]
and this is the largest closed interval contained in every member of
$\mathbf{J}$.
\end{proposition}

\begin{definition}
\label{cnt0211}Given $A,B\in\mathbf{I}$, define $\mathbf{S}(A,B)\mathbf{\doteq
}$ $\mathbf{\{}C$: $C\in\mathbf{I}$, $A\subseteq C,B\subseteq C\mathbf{\}}$.
Then we define
\[
A\overset{.}{\cup}B\doteq\cap_{C\in\mathbf{S}(A,B)}C.
\]
\end{definition}

\begin{proposition}
\label{cnt0212}The structure $(\mathbf{I},\subseteq,\overset{.}{\cup},\cap)$
is a lattice with respect to the $\subseteq$ order (i.e. set theoretic
inclusion). Given any intervals $A=[a_{1},a_{2}]$ $\in\mathbf{I}$,
$B=[b_{1},b_{2}]\in\mathbf{I}$, $\sup(A,B)$ = $A\overset{.}{\cup}B$ =
$[a_{1}\sqcap b_{1},a_{2}\sqcup b_{2}]$, \ $\inf(A,B)$= $A\cap$ $B$ =
$[a_{1}\sqcup b_{1}$ , $a_{2}\sqcap b_{2}]$.
\end{proposition}

\textbf{Remark}. In other words, given any intervals $A=[a_{1},a_{2}]$,
$B=[b_{1},b_{2}]$, $[a_{1}\sqcap b_{1},a_{2}\sqcup b_{2}]$ is the smallest
closed interval which contains both $A$ and $B\ $and [$a_{1}\sqcup b_{1}$ ,
$a_{2}\sqcap b_{2}$] is the largest closed interval contained in both $A$ and
$B$.

We define \emph{fuzzy sublattices} and \emph{fuzzy convex sublattices} in
terms of their $p$-cuts; this is different from, but equivalent to Ajmal's
approach \cite{AjLat01}.

\begin{definition}
\label{cnt0213}We say $M:X\rightarrow L$ is a \emph{fuzzy sublattice} of
$(X,\sqsubseteq)$ iff $\forall p\in L$ the set$\ M_{p}$ is a sublattice of
$(X,\sqsubseteq).$
\end{definition}

\begin{definition}
\label{cnt0214}We say $M:X\rightarrow L$ is a \emph{fuzzy convex sublattice}
of $(X,\sqsubseteq)$ iff $\forall p\in L$ the set$\ M_{p}$ is a convex
sublattice of $(X,\sqsubseteq)$; (i.e. $\forall p\in L,\forall x,y\in M_{p}$
we have $[x\sqcap y,x\sqcup y]\subseteq M_{p}$).
\end{definition}

\begin{proposition}
\label{cnt0215}$M:X\rightarrow L$ is a fuzzy sublattice of $(X,\sqsubseteq)$
iff
\[
\forall x,y\in X:\quad M(x\sqcap y)\wedge M(x\sqcup y)\geq M(x)\wedge M(y).
\]
\end{proposition}

\begin{proof}
See \cite{Tep01}.
\end{proof}

\begin{proposition}
\label{cnt0216}Let $M:X\rightarrow L$ be a fuzzy sublattice of $(X,\sqsubseteq
)$. It is a fuzzy convex sublattice of $(X,\sqsubseteq)$ iff
\begin{equation}
\forall x,y\in X,\forall z\in\lbrack x\sqcap y,x\sqcup y]:\quad M(z)\geq
M(x\sqcap y)\wedge M(x\sqcup y)=M(x)\wedge M(y). \label{Eq02}%
\end{equation}
\end{proposition}

\begin{proof}
(i)\ Assume $M$ is a fuzzy convex sublattice. Choose any $x,y\in X$. Set
$p_{1}=M(x\sqcap y)$, $p_{2}=M(x\sqcup y)$; then $x\sqcap y$, $x\sqcup y\in$
$M_{p_{1}\wedge p_{2}}$. Take any $z\in\lbrack x\sqcap y,x\sqcup y]$. Since
$M$ is a fuzzy convex sublattice: $z\in M_{p_{1}\wedge p_{2}}\Rightarrow$
$M(z)\geq p_{1}\wedge p_{2}$ = $M(x\sqcap y)\wedge M(x\sqcup y)$. Since
$x,y\in$\ $[x\sqcap y,x\sqcup y]$ we have $M(x)\geq M(x\sqcap y)\wedge
M(x\sqcup y)$, $M(y)\geq M(x\sqcap y)\wedge M(x\sqcup y)$; and so $M(x)\wedge
M(y)\geq M(x\sqcap y)\wedge M(x\sqcup y)$. On the other hand, since $M$ is a
fuzzy sublattice, from Proposition \ref{cnt0215} we have $M(x\sqcap y)\wedge
M(x\sqcup y)\geq$ $M(x)\wedge M(y)$. Hence $M(x\sqcap y)\wedge M(x\sqcup
y)\ \ $= $M(x)\wedge M(y)$.

(ii)\ Conversely, assume (\ref{Eq02}) holds. Take any $p\in L$. If $M_{p}$ is
empty, \ then it is a convex sublattice. If $M_{p}$ is not empty, take any
$x,y\in M_{p}$. Set $p_{1}=M(x)$, $p_{2}=M(y)$. We have $x\in M_{p}%
\Rightarrow$ $p_{1}=M(x)\geq p$, $y\in M_{p}\Rightarrow$ $p_{2}=M(y)\geq p$.
From (\ref{Eq02}) we have $M(x\sqcap y)\geq M(x)\wedge M(y)=p_{1}\wedge
p_{2}\geq p\Rightarrow$ $x\sqcap y\in M_{p}$. Similarly $x\sqcup y\in M_{p}$
\ and so $M_{p}$ is a sublattice. Set $q_{1}=M(x\sqcap y)$, $q_{2}=M(x\sqcup
y)$. Now take any $z\in\lbrack x\sqcap y,x\sqcup y]$. From (\ref{Eq02}) we
have $M(z)\geq q_{1}\wedge q_{2}=p_{1}\wedge p_{2}\geq p\Rightarrow$ $z\in
M_{p}$. Hence $M_{p}$ is a convex sublattice for all $p\in L$, i.e. $M\ $ is a
fuzzy convex sublattice.
\end{proof}

\section{The Lattice of Fuzzy Intervals}

\label{sec03}

We now introduce \emph{fuzzy intervals}.

\begin{definition}
\label{cnt0301}We say $M:X\rightarrow L$ is a \emph{fuzzy interval }\ of
$(X,\sqsubseteq)$ iff
\[
\forall p\in L:\quad M_{p}\text{ is a closed interval of }(X,\leq).
\]
The collection all fuzzy intervals will be denoted by $\widetilde{\mathbf{I}%
}(X,L)$ or simply by $\widetilde{\mathbf{I}}$.
\end{definition}

The following proposition will be often used in the sequel. It states that an
arbitrary intersection of fuzzy intervals yields a fuzzy interval.

\begin{proposition}
\label{cnt0302}For all $\widetilde{\mathbf{J}}\subseteq\widetilde{\mathbf{I}}$
we have: $\wedge_{M\in\widetilde{\mathbf{J}}}M\in\widetilde{\mathbf{I}} $
\end{proposition}

\begin{proof}
Choose any $\widetilde{\mathbf{J}}\subseteq\widetilde{\mathbf{I}}%
\subseteq\mathbf{F}$. The fuzzy set $\wedge_{M\in\widetilde{\mathbf{J}}}M$ is
well defined, in view of the fact that $(\mathbf{F},\leq,\vee,\wedge)$ is a
complete lattice. Choose any $p\in L$. It is easy to show that $(\wedge
_{M\in\widetilde{\mathbf{J}}}M)_{p}$ = $\cap_{M\in\widetilde{\mathbf{J}}}%
M_{p}$. Then for every $M\in\widetilde{\mathbf{J}}$ , the cut $M_{p}$ will be
a closed interval (perhaps the empty interval). From Proposition
\ref{cnt0210}, an arbitrary intersection of closed intervals yields a closed
interval. Hence, for every $p\in L$ the set $(\wedge_{M\in\widetilde
{\mathbf{J}}}M)_{p}$ is a closed interval, i.e. $\wedge_{M\in\widetilde
{\mathbf{J}}}M$ is a fuzzy interval.
\end{proof}

Since $\widetilde{\mathbf{I}}\subseteq\mathbf{F}$, it follows that
$(\widetilde{\mathbf{I}},\leq)\ $is an ordered set. We now establish (using
Proposition \ref{cnt0302}) that $(\widetilde{\mathbf{I}},\leq)$ is a lattice.

\begin{definition}
\label{cnt0303}For all $M,N\in\widetilde{\mathbf{I}}$ we define $M\overset
{.}{\vee}N$ as follows. We define $\widetilde{\mathbf{S}}(M,N)\doteq$
$\{A:A\in\widetilde{\mathbf{I}}$, $M\leq A,N\leq A\}$ and then define
\[
M\overset{.}{\vee}N\doteq\wedge_{A\in\widetilde{\mathbf{S}}(M,N)}A.
\]
\end{definition}

\begin{proposition}
\label{cnt0304}$(\widetilde{\mathbf{I}},\leq,\overset{.}{\vee},\wedge)\ $is a
complete lattice.
\end{proposition}

\begin{proof}
(i)$\;M\wedge N$ is the infimum in $\mathbf{F}$ of $M$ and $N$. From
Proposition \ref{cnt0302} we have $M\wedge N\in\widetilde{\mathbf{I}}$, hence
$M\wedge N$ is also the infimum of $M$ and $N$ in $\widetilde{\mathbf{I}}$.

(ii)\ For all $A\in\widetilde{\mathbf{S}}(M,N)$ we have $M\leq A$ and so
$M\leq\wedge_{A\in\widetilde{\mathbf{S}}(M,N)}A=M\overset{.}{\vee}N$;
similarly $N\leq M\overset{.}{\vee}N$. Furthermore, if there is some
$B\in\widetilde{\mathbf{I}}$ such that $M\leq B$, $N\leq B$, then
$B\in\widetilde{\mathbf{S}}(M,N)$. Hence $M\overset{.}{\vee}N$ = $\wedge
_{A\in\widetilde{\mathbf{S}}(M,N)}A\leq B$. \ Finally, since $\widetilde
{\mathbf{S}}(M,N)$ $\subseteq\widetilde{\mathbf{I}}$, we have $M\overset
{.}{\vee}N$ = $\wedge_{A\in\widetilde{\mathbf{S}}(M,N)}A\in\widetilde
{\mathbf{I}}$. Hence $M\overset{.}{\vee}N$ is the supremum in $\widetilde
{\mathbf{I}}$ of $M$ and $N$.\noindent

(iii) To establish completeness of $(\widetilde{\mathbf{I}},\leq,\overset
{.}{\vee},\wedge)\ $we must show that any $\widetilde{\mathbf{J}}%
\subseteq\widetilde{\mathbf{I}}$ has an infimum and a supremum in
$\widetilde{\mathbf{I}}$. We have already remarked (Proposition \ref{cnt0302})
that, for any $\widetilde{\mathbf{J}}\subseteq\widetilde{\mathbf{I}}$, the set
$\wedge_{M\in\widetilde{\mathbf{J}}}M$ is a well defined fuzzy interval. Since
$\wedge\widetilde{\mathbf{J}}$ $\ $ = $\wedge_{M\in\widetilde{\mathbf{J}}}M $
is the infimum of $\widetilde{\mathbf{J}}$ in $\mathbf{F}$, it will also be
the infimum of $\widetilde{\mathbf{J}}$ in $\widetilde{\mathbf{I}}%
\subseteq\mathbf{F}$. Regarding the supremum, we must define appropriately
$\overset{.}{\vee}\widetilde{\mathbf{J}}$. Define a set $\widetilde
{\mathbf{S}}(\widetilde{\mathbf{J}})=\{A\in\widetilde{\mathbf{I}}:$ $\forall
M\in\widetilde{\mathbf{J}}$ we have $\ M\leq A\}$. Define $\overset{.}{\vee
}\widetilde{\mathbf{J}}\doteq\wedge_{A\in\widetilde{\mathbf{S}}(\widetilde
{\mathbf{J}})}A$. Then $\overset{.}{\vee}\widetilde{\mathbf{J}}\in
\widetilde{\mathbf{I}}$ (as an intersection of fuzzy intervals), and it is
easy to show that: $\forall M\in\widetilde{\mathbf{J}}$ we have $M\leq
\overset{.}{\vee}\widetilde{\mathbf{J}}$, $\forall A\in\widetilde{\mathbf{S}%
}(\widetilde{\mathbf{J}})$ we have $\overset{.}{\vee}\widetilde{\mathbf{J}%
}\leq A$. Hence $\overset{.}{\vee}\widetilde{\mathbf{J}}$ is the supremum of
$\widetilde{\mathbf{J}}$ and completeness has been established.
\end{proof}

The following propositions establish some properties of fuzzy intervals.

\begin{definition}
\label{cnt0305}For every fuzzy set $M$ we define $L_{M}\doteq\{p:M_{p}%
\neq\emptyset\}$.
\end{definition}

\begin{proposition}
\label{cnt0306}(i)\ Let $M$ be a fuzzy convex sublattice. If we have
\begin{equation}
\forall p\in L_{M}:M(\sqcap M_{p})\geq\wedge_{x\in M_{p}}M(x),\quad M(\sqcup
M_{p})\geq\wedge_{x\in M_{p}}M(x), \label{Eq31}%
\end{equation}
then $M$ is a fuzzy interval.

(ii) If $M$ is a fuzzy interval, then it is a fuzzy convex sublattice and we
have
\[
\forall p\in L_{M}:M(\sqcap M_{p})\geq\wedge_{x\in M_{p}}M(x),\quad M(\sqcup
M_{p})\geq\wedge_{x\in M_{p}}M(x).
\]
\end{proposition}

\begin{proof}
(i) Assume (\ref{Eq31}) holds. Choose any $p\in L_{M}$. Now, by completenes of
$(X,\sqsubseteq)$, $\sqcap M_{p}$ and $\sqcup M_{p}$ exist. Clearly
$M_{p}\subseteq\lbrack\sqcap M_{p},\sqcup M_{p}]$. On the other hand, from
(\ref{Eq31}), $M(\sqcap M_{p})\geq\wedge_{x\in M_{p}}M(x)\geq p\Rightarrow$
$\sqcap M_{p}\in M_{p}$, i.e. $M_{p}$ contains its infimum. Similarly
$M(\sqcup M_{p})\geq\wedge_{x\in M_{p}}M(x)\geq p\Rightarrow$ $\sqcup M_{p}\in
M_{p}$. Since $M_{p}$ is a convex sublattice and $\sqcap M_{p},\sqcup M_{p}\in
M_{p}$, it follows that $[\sqcap M_{p},\sqcup M_{p}]\subseteq M_{p}$. Hence
for all $p\in L_{M}$ we have that $M_{p}=[\sqcap M_{p},\sqcup M_{p}]$.
Further, for all $p\in L-L_{M}$, $M_{p}$ is the empty set, which is considered
a closed interval. Hence for all $p\in L$ the set $M_{p}$ is a closed
interval, i.e. $M$ is a fuzzy interval.

(ii) If $M$ is a fuzzy interval then for all $p\in L_{M}$ we have $M_{p}$ =
[$\sqcap M_{p}$, $\sqcup M_{p}$], which is a closed interval and \emph{a
fortiori} a convex sublattice. Hence$\;M$ is a fuzzy convex sublattice.
Furthermore, $M_{p}$ = [$\sqcap M_{p}$, $\sqcup M_{p}$]$\Rightarrow$ $\sqcap
M_{p}\in M_{p}$ $\Rightarrow$ $M(\sqcap M_{p})$ $\geq$ $\wedge_{x\in M_{p}%
}M(x)$. Similarly, $\sqcup M_{p}\in M_{p}$ $\Rightarrow$ $M(\sqcup M_{p})$
$\geq$ $\wedge_{x\in M_{p}}M(x)$ .
\end{proof}

\begin{corollary}
\label{cnt0307}If $M$ is a fuzzy interval, then $\forall p\in L_{M}$ we have
$M(\sqcap M_{p})\wedge M(\sqcup M_{p})$ = $\wedge_{x\in M_{p}}M(x).$
\end{corollary}

\begin{corollary}
\label{cnt0308}Let $X$ be finite. Then every fuzzy convex sublattice is a
fuzzy interval and conversely.
\end{corollary}

\begin{proposition}
\label{cnt0309}If $M$ is a fuzzy interval, then $\forall p\in L_{M}$ we have
$M_{p}$ = $M_{p_{1}\wedge p_{2}}$, where $p_{1}=M(\sqcap M_{p})$,
$p_{2}=M(\sqcup M_{p}).$
\end{proposition}

\begin{proof}
Choose any $p\in L_{M}$. Since $M$ is a fuzzy interval, we have $M_{p}=[\sqcap
M_{p},\sqcup M_{p}].\ $Set $p_{1}=M(\sqcap M_{p})\geq p$,\ $p_{2}=M(\sqcup
M_{p})\geq p$. Then $M(\sqcap M_{p})=p_{1}\geq p_{1}\wedge p_{2}$ and so
$\sqcap M_{p}\in M_{p_{1}\wedge p_{2}}$. Similarly $\sqcup M_{p}\in
M_{p_{1}\wedge p_{2}}$. Since $M$ is a fuzzy interval (and so a fuzzy convex
sublattice)\ it follows that $[\sqcap M_{p},\sqcup M_{p}]\subseteq
M_{p_{1}\wedge p_{2}}$. On the other hand $p_{1}\wedge p_{2}\geq p\Rightarrow$
$M_{p_{1}\wedge p_{2}}\subseteq M_{p}$ = $[\sqcap M_{p},\sqcup M_{p}]$. Hence
$M_{p_{1}\wedge p_{2}}$ = $M_{p}$.
\end{proof}

\section{Distributivity}

\label{sec04}

In all of this section we assume $(X,\sqsubseteq,\sqcup,\sqcap)$ to be
\emph{completely distributive} according to the following definition.

\begin{definition}
\label{cnt0401}The lattice $(X,\sqsubseteq,\sqcup,\sqcap)$ is said to be
\emph{completely distributive}, iff for every set $Y\subseteq X$ we have
$x\sqcup(\sqcap_{y\in Y}y)=\sqcap_{y\in Y}(x\sqcup y),\qquad x\sqcap
(\sqcup_{y\in Y}y)=\sqcup_{y\in Y}(x\sqcap y).$
\end{definition}

Let $M,N$ be fuzzy intervals. Our first task is to establish some properties
of the cuts $\left(  M\wedge N\right)  _{p}\ $\ and $(M\overset{.}{\vee}N)_{p}
$. From Proposition \ref{cnt0304} we see that $M\wedge N\ $\ and $M\overset
{.}{\vee}N$ are fuzzy intervals; hence $\forall p\in L$ the cuts $\left(
M\wedge N\right)  _{p}\ $\ and $(M\overset{.}{\vee}N)_{p}$ are (crisp) closed intervals.

\begin{definition}
\label{cnt0402}For all $M,N\in\widetilde{\mathbf{I}}$ and for all $p\in L$ we
define $C_{p}(M,N)$ $=M_{p}\cap N_{p}$.
\end{definition}

\begin{proposition}
\label{cnt0403}For all $M,N\in\widetilde{\mathbf{I}}$ and for all $p\in L$ we
have: $(M\wedge N)_{p}=C_{p}(M,N\mathbf{)}$.
\end{proposition}

\begin{proof}
Take any $M,N\in\widetilde{\mathbf{I}}$, any $p\in L$. We have $x\in(M\wedge
N)_{p}$ $\Leftrightarrow$ $\left(  M\wedge N\right)  (x)\geq p$
$\Leftrightarrow$ $M(x)\wedge N(x)\geq p$ $\Leftrightarrow$ $\left(  M(x)\geq
p\text{ and }N(x)\geq p\right)  $ $\Leftrightarrow$ $\left(  x\in M_{p}\text{
and }x\in N_{p}\right)  $ $\Leftrightarrow$ $x\in M_{p}\cap N_{p}%
=C_{p}(M,N\mathbf{)}$.
\end{proof}

\begin{proposition}
\label{cnt0404}Take any $M,N\in\widetilde{\mathbf{I}}$. We have:

(i)\ $\forall p,q\in L$ :\ $p\leq q\Rightarrow$ $C_{q}(M,N)$ $\subseteq
C_{p}(M,N)$,

(ii) $\forall P\subseteq L$: $\cap_{p\in P}C_{p}(M,N)$ = $C_{\vee P}(M,N)$.

(iii) $C_{0}(M,N)=X$.
\end{proposition}

\begin{proof}
These properties follow from the fact that for all $p\in L$ we have
$C_{p}(M,N\mathbf{)}$ = $(M\wedge N)_{p}$, i.e. the family $\{C_{p}%
(M,N\mathbf{)\}}_{p\in L}$ is a family of cuts.
\end{proof}

Hence we have characterized the cuts of $M\wedge N$ in terms of the cuts of
$M$ and $N$. We will now do the same for the cuts of $M\overset{.}{\vee}N$.
However, before proceeding we need some auxiliary definitions and propositions.

\begin{definition}
\label{cnt0405}For every $M\in\widetilde{\mathbf{I}}$, we define the functions
$\underline{M}:L\rightarrow X$, $\overline{M}:L\rightarrow X$ as follows. For
$p\in L_{M}$, $\underline{M}(p)\doteq\sqcap M_{p}$, $\overline{M}%
(p)\doteq\sqcup M_{p}$; for $p\in L-L_{M}$, $\underline{M}(p)\doteq\sqcup X$,
$\overline{M}(p)\doteq\sqcap X$.
\end{definition}

\textbf{Remark}. Hence we can write $M_{p}=[\underline{M}(p),\overline{M}(p)]
$ for \emph{every }$p\in L$. Because: if $p\in L_{M}$, then $M_{p}$ = [$\sqcap
M_{p},\sqcup M_{p}$] = $[\underline{M}(p),\overline{M}(p)]$; if $p\in L-L_{M}%
$, then $M_{p}=\emptyset$ = [$\sqcup X,\sqcap X$] = $[\underline
{M}(p),\overline{M}(p)]$.

\begin{proposition}
\label{cnt0406}Take any $M\in\widetilde{\mathbf{I}}$ and for all $p\in L$ set
$M_{p}=[\underline{M}(p),\overline{M}(p)]$. Then

(i) $\forall p,q\in L:p\leq q\Rightarrow\left(  \underline{M}(p)\sqsubseteq
\underline{M}(q),\overline{M}(p)\sqsupseteq\overline{M}(q)\right)  $ .

(ii)\ $\forall P\subseteq L:\sqcup_{p\in P}\underline{M}(p)=\underline{M}(\vee
P)$, $\sqcap_{p\in P}\overline{M}(p)=\overline{M}(\vee P)$.
\end{proposition}

\begin{proof}
(i)\ Since $\left\{  M_{p}\right\}  _{p\in P}$ are cuts, from
Prop.\ref{cnt0206}.(i) we have: $p\leq q\Rightarrow M_{q}\subseteq
M_{p}\Rightarrow$ [$\underline{M}(q)$ , $\overline{M}(q)$] $\subseteq$
[$\underline{M}(p)$ , $\overline{M}(p)$]$\Rightarrow$ ($\underline{M}%
(p)\leq\underline{M}(q)$ , $\overline{M}(p)\geq\overline{M}(q)$). Note in
particular that: if $q\notin L_{M}$, then $\underline{M}(p)\sqsubseteq
\underline{M}(q)=\sqcup X$ and $\overline{M}(p)\sqsupseteq\overline
{M}(q)=\sqcap X$.

(ii) Since $\left\{  M_{p}\right\}  _{p\in P}$ are cuts, from
Prop.\ref{cnt0206}.(ii) we have:\ $\cap_{p\in P}M_{p}=M_{\vee P}$. But
$M_{\vee P}$\ = [\underline{$M$}$(\vee P)$ , $\overline{M}(\vee P)$] and
(Proposition \ref{cnt0210})\ $\cap_{p\in P}M_{p}$= [$\sqcup_{p\in P}%
\underline{M}(p)$ , $\sqcap_{p\in P}\overline{M}(p)$] which yields the
required result. Note in particular that:\ if there exists some $q\in P$ such
that $q\in L-L_{M}$, then $M_{q}=\emptyset$, \ $\cap_{p\in P}M_{p}=\emptyset$,
and $M_{\vee P}=\emptyset$ = $[\underline{M}(\vee P),\overline{M}(\vee P)]$
with $\underline{M}(\vee P)$ = $\sqcup X$, $\overline{M}(\vee P)$ = $\sqcap
X$. Also, in this case $\underline{M}(q)=\sqcup X$, $\sqcup_{p\in P}%
\underline{M}(p)=\sqcup X$, $\overline{M}(q)=\sqcap X$, $\sqcap_{p\in
P}\underline{M}(p)=\sqcap X$.
\end{proof}

\begin{proposition}
\label{cnt0407}(i) Take any $P\subseteq L$ and any functions $F:L\rightarrow
X$, $G:L\rightarrow X$ which satisfy
\begin{align*}
p  &  \leq q\Rightarrow F(p)\sqsubseteq F(q),\qquad\sqcup_{p\in P}F(p)=F(\vee
P),\\
p  &  \leq q\Rightarrow G(p)\sqsubseteq G(q),\qquad\sqcup_{p\in P}G(p)=G(\vee
P).
\end{align*}
Then $\sqcup_{p\in P}\left(  F(p)\sqcap G(p)\right)  =F(\vee P)\sqcap G(\vee
P)$.

(ii) Take any $P\subseteq L$ and any functions $F:L\rightarrow X$,
$G:L\rightarrow X$ which satisfy
\begin{align*}
p  &  \leq q\Rightarrow F(p)\sqsupseteq F(q),\qquad\sqcap_{p\in P}F(p)=F(\vee
P),\\
p  &  \leq q\Rightarrow G(p)\sqsupseteq G(q),\qquad\sqcap_{p\in P}G(p)=G(\vee
P).
\end{align*}
Then $\sqcap_{p\in P}\left(  F(p)\sqcup G(p)\right)  =F(\vee P)\sqcup G(\vee
P)$.
\end{proposition}

\begin{proof}
For (i), take any $p\in P$. Then $F(p)\sqcap G(p)\sqsubseteq F(p).$ Hence
$\sqcup_{p\in P}\left(  F(p)\sqcap G(p)\right)  \sqsubseteq\sqcup_{p\in
P}F(p)$ = $F(\vee P)$. Similarly $\sqcup_{p\in P}\left(  F(p)\sqcap
G(p)\right)  \sqsubseteq\sqcup_{p\in P}G(p)$ = $G(\vee P)$. It follows that
\begin{equation}
\sqcup_{p\in P}\left(  F(p)\sqcap G(p)\right)  \sqsubseteq F(\vee P)\sqcap
G(\vee P). \label{Eq41}%
\end{equation}
On the other hand, using complete distributivity, we have $\sqcup_{p\in P,q\in
P}\left(  F(p)\sqcap G(q)\right)  $ = $\sqcup_{p\in P}\left(  F(p)\sqcap
\left(  \sqcup_{q\in P}G(q)\right)  \right)  $ = $\sqcup_{p\in P}\left(
F(p)\sqcap G(\vee P)\right)  $ = $\left(  \sqcup_{p\in P}F(p)\right)  \sqcap
G(\vee P)$ = $F(\vee P)\sqcap G(\vee P)$. In short
\begin{equation}
F(\vee P)\sqcap G(\vee P)=\sqcup_{p\in P,q\in P}\left(  F(p)\sqcap
G(q)\right)  \label{Eq42}%
\end{equation}
Finally, since $(L,\leq)$ is totally ordered, $P$ is a sublattice of
$(L,\leq);$ so for any $p,q\in P$ we have $p\vee q\in P$. Then $\left(  p\leq
p\vee q\text{, }q\leq p\vee q\right)  \Rightarrow$ $F(p)\sqcap G(q)\sqsubseteq
$ $F(p\vee q)\sqcap G(p\vee q)$. So $\sqcup_{p\in P,q\in P}\left(  F(p)\sqcap
G(q)\right)  \sqsubseteq$ $\sqcup_{p\in P,q\in P}\left(  F(p\vee q)\sqcap
G(p\vee q)\right)  \sqsubseteq$ $\sqcup_{r\in P}\left(  F(r)\sqcap
G(r)\right)  $. Hence
\begin{equation}
\sqcup_{p\in P,q\in P}\left(  F(p)\sqcap G(q)\right)  \sqsubseteq\sqcup_{p\in
P}\left(  F(p)\sqcap G(p)\right)  \label{Eq43}%
\end{equation}
From (\ref{Eq41}), (\ref{Eq42}), (\ref{Eq43}) follows that $\sqcup_{p\in
P}\left(  F(p)\sqcap G(p)\right)  $ = $F(\vee P)\sqcap G(\vee P)$ and (i)
\ has been proved; (ii)\ is proved dually.
\end{proof}

Now we return to the cuts of $M\overset{.}{\vee}N$.

\begin{definition}
\label{cnt0408}For all $M,N\in\widetilde{\mathbf{I}}$ and for all $p\in L$ we
define $D_{p}(M,N)=M_{p}\overset{.}{\cup}N_{p}$.
\end{definition}

\begin{proposition}
\label{cnt0409}Take any $M,N\in\widetilde{\mathbf{I}}$ . We have

(i) $\forall p,q\in L$:\ $p\leq q\Rightarrow$ $D_{q}(M,N)$ $\subseteq
D_{p}(M,N)$,

(ii)$\;\forall P\subseteq L$ : $\cap_{p\in P}D_{p}(M,N)$ = $D_{\vee P}(M,N)$.

(iii) $D_{0}(M,N)=X$.
\end{proposition}

\begin{proof}
(i) Assume $p\leq q$. Then $\left(  M_{q}\subseteq M_{p},N_{q}\subseteq
N_{p}\right)  \Rightarrow$ $M_{q}\overset{.}{\cup}N_{q}\subseteq M_{p}%
\overset{.}{\cup}N_{p}\Rightarrow$ $D_{q}(M,N)$ $\subseteq D_{p}(M,N)$.

(ii)\ Take any $P\subseteq L$ and any $p\in P$. We have $D_{p}%
(M,N)=[\underline{M}(p)\sqcap\underline{N}(p),\overline{M}(p)\sqcup
\overline{N}(p)]$, hence
\begin{equation}
\cap_{p\in P}D_{p}(M,N)=[\sqcup_{p\in P}(\underline{M}(p)\sqcap\underline
{N}(p)),\sqcap_{p\in P}(\overline{M}(p)\sqcup\overline{N}(p))]. \label{Eq44}%
\end{equation}
Also
\begin{equation}
D_{\vee P}(M,N)=[\underline{M}(\vee P)\sqcap\underline{N}(\vee P),\overline
{M}(\vee P)\sqcup\overline{N}(\vee P)]. \label{Eq45}%
\end{equation}
Use Proposition \ref{cnt0407}.(i) with $F(p)=\underline{M}(p)$ and
$G(p)=\underline{N}(p)$. Then
\begin{equation}
\sqcup_{p\in P}(\underline{M}(p)\sqcap\underline{N}(p))=\underline{M}(\vee
P)\sqcap\underline{N}(\vee P). \label{Eq46}%
\end{equation}
Use Proposition \ref{cnt0407}.(ii) with $F(p)=\overline{M}(p)$ and
$G(p)=\overline{N}(p)$. Then
\begin{equation}
\sqcap_{p\in P}(\overline{M}(p)\sqcup\overline{N}(p))=\overline{M}(\vee
P)\sqcup\overline{N}(\vee P). \label{Eq47}%
\end{equation}
Eqs.(\ref{Eq44}--\ref{Eq47})\ yield the required result.

(iii)\ $D_{0}(M,N)=M_{0}\overset{.}{\cup}N_{0}=X\overset{.}{\cup}X=X$.
\end{proof}

\begin{proposition}
\label{cnt0410}For all $M,N\in\widetilde{\mathbf{I}}$ and for all $p\in L$ we
have: $(M\overset{.}{\vee}N)_{p}=D_{p}(M,N\mathbf{)}$
\end{proposition}

\begin{proof}
From Proposition \ref{cnt0409} follows that $\left\{  D_{p}(M,N)\right\}
_{p\in L}$ is a family of cuts. Hence, if we define a fuzzy set $(M\veebar N)
$ by setting
\[
\forall x\in X:(M\veebar N)(x)\doteq\vee\left\{  p:x\in D_{p}(M,N)\right\}
\]
then $\forall p\in L$ we will have $(M\veebar N)_{p}=D_{p}(M,N)$ (Proposition
\ref{cnt0207}). From this also follows that $(M\veebar N)$ is a fuzzy interval
(since $\forall p\in L$ we have $(M\veebar N)_{p}$ = $D_{p}(M,N)$ =
$M_{p}\overset{.}{\cup}N_{p}$). Now choose any $p\in L$ ; we will show that
$(M\overset{.}{\vee}N)_{p}=(M\veebar N)_{p}$.

First, $(M\overset{.}{\vee}N)_{p}$ is a (crisp) closed interval. Also, $x\in
M_{p}\Rightarrow$ $(M\overset{.}{\vee}N)(x)\geq M(x)\geq p\ \Rightarrow$
$x\in(M\overset{.}{\vee}N)_{p}$. So $M_{p}\subseteq(M\overset{.}{\vee}N)_{p}$.
Similarly $N_{p}\subseteq(M\overset{.}{\vee}N)_{p}$. Hence $(M\overset{.}%
{\vee}N)_{p}\in\mathbf{S}(M_{p},N_{p})$ which implies that $(M\veebar
N)_{p}=D_{p}(M,N)$ = $M_{p}\overset{.}{\cup}N_{p}$ = $\cap_{A\in
\mathbf{S}(M_{p},N_{p})}A\subseteq$ $(M\overset{.}{\vee}N)_{p}$.

Second, choose any $x\in X$ and set $p=M(x)$. Then $x\in M_{p}\subseteq
D_{p}(M,N)=(M\veebar N)_{p}$. Hence $(M\veebar N)(x)\geq p=M(x)$; similarly
$(M\veebar N)(x)\geq N(x)$. Since $M\overset{.}{\vee}N=\sup(M,N)$, it follows
that $(M\veebar N)(x)\geq(M\overset{.}{\vee}N)(x)$ and so $(M\veebar
N)_{p}\supseteq(M\overset{.}{\vee}N)_{p}$.

So we have $(M\veebar N)_{p}=(M\overset{.}{\vee}N)_{p}$ which (Proposition
\ref{cnt0205}) implies $M\veebar N=M\overset{.}{\vee}N$.\noindent
\end{proof}

\begin{proposition}
\label{cnt0411}$(\widetilde{\mathbf{I}},\leq,\overset{.}{\vee},\wedge)\ $is a
distributive lattice.
\end{proposition}

\begin{proof}
We must show that for any $A,B,C\in\widetilde{\mathbf{I}}$ we have
$(A\overset{.}{\vee}B)\wedge C$ = $(A\wedge C)\overset{.}{\vee}(B\wedge C)$
and $(A\wedge B)\overset{.}{\vee}C$ = $(A\overset{.}{\vee}C)\wedge
(B\overset{.}{\vee}C)$. We will show this by showing equality of the $p$-cuts.

Indeed, choose any $p\in L$ and set $A_{p}=[a_{1},a_{2}]$, $B_{p}=[b_{1}%
,b_{2}]$, $C_{p}=[c_{1},c_{2}]$ (in case any of these intervals is empty,
denote it by $[\sqcap X,\sqcup X]$). Now
\[
\left(  (A\overset{.}{\vee}B)\wedge C\right)  _{p}=(A\overset{.}{\vee}%
B)_{p}\cap C_{p}=(A_{p}\overset{.}{\cup}B_{p})\cap C_{p}=
\]%
\[
([a_{1},a_{2}]\overset{.}{\cup}[b_{1},b_{2}])\cap\lbrack c_{1},c_{2}%
]=[a_{1}\sqcap b_{1},a_{2}\sqcup b_{2}]\cap\lbrack c_{1},c_{2}]=
\]%
\[
\lbrack(a_{1}\sqcap b_{1})\sqcup c_{1},(a_{2}\sqcup b_{2})\sqcap
c_{2}]=[(a_{1}\sqcup c_{1})\sqcap(b_{1}\sqcup c_{1}),(a_{2}\sqcap c_{2}%
)\sqcup(b_{2}\sqcap c_{2}))]=
\]%
\[
\lbrack a_{1}\sqcup c_{1},a_{2}\sqcap c_{2}]\overset{.}{\cup}[b_{1}\sqcup
c_{1},b_{2}\sqcap c_{2}]\ =\left(  [a_{1},a_{2}]\cap\lbrack c_{1}%
,c_{2}]\right)  \overset{.}{\cup}\left(  [b_{1},b_{2}]\cap\lbrack c_{1}%
,c_{2}]\right)  \ =
\]%
\[
(A_{p}\cap C_{p})\overset{.}{\cup}(B_{p}\cap C_{p})=(A\wedge C)_{p}\overset
{.}{\cup}(B\wedge C)_{p}=\left(  (A\wedge C)\overset{.}{\vee}(B\wedge
C)\right)  _{p}.
\]
Since for all $p\in L$ we have $\left(  (A\overset{.}{\vee}B)\wedge C\right)
_{p}$ = $\left(  (A\wedge C)\overset{.}{\vee}(B\wedge C)\right)  _{p}$ , it
follows that $(A\overset{.}{\vee}B)\wedge C$ = $(A\wedge C)\overset{.}{\vee
}(B\wedge C)$. Dually we show that $(A\wedge B)\overset{.}{\vee}C$ =
$(A\overset{.}{\vee}C)\wedge(B\overset{.}{\vee}C)$.
\end{proof}

\section{Discussion}

\label{sec05}

In this paper we have introduced fuzzy intervals and obtained some of their
basic properties. The method we have used is rather standard in the study of
fuzzy algebras -- in particular we have obtained several properties of fuzzy
intervals by studying their $p$-cuts. This method can be used to obtain
further properties of fuzzy intervals.

In our analysis we have made several assumptions, the most prominent ones
being that:\ (a)\ $L$ is $[0,1]$ and (b)$\;X$ is complete and completely
distributive. To what extent can these assumptions be relaxed?

Regarding $L$, the analysis remains unchanged if $(L,\leq,\vee,\wedge)$ is
simply a chain. But it does not seem obvious how to generalize our results to
L-fuzzy lattices, because Proposition \ref{cnt0407}\ requires that for every
$P\subseteq L$, and for all $p,q\in P$, we have $p\vee q\in P$; for this to be
true for arbitrary $P\subseteq L$, \ $(L,\leq)$ must be a chain.

The completeness of $(X,\sqsubseteq,\sqcup,\sqcap)$ is also essential.
Obviously, if $(X,\sqsubseteq,\sqcup,\sqcap)$ is not complete, there is no
guarantee that an infinite union of fuzzy intervals will be a fuzzy interval.
Regarding complete distributivity, it has only been used in Section
\ref{sec04}, but there it plays an essential role in the proof of Proposition
\ref{cnt0407}. Let us note that in the important special case where $X$ has
finite cardinality, completeness is automatically satisfied and complete
distributivity is equivalent to distributivity (which clearly is a minimum
requirement for the lattice of fuzzy intervals to be distributive).

Finally, let us discuss briefly the important special case when
$(X,\sqsubseteq,\sqcup,\sqcap)$ = $(R,\leq,\vee,\wedge)$. In this case we
obtain the ``classical'' notion of a fuzzy interval, i.e. a fuzzy set such
that its $p$-cuts are closed intervals on the real line (compare \cite[p.37,
p.48]{Nguyen}. It is worth noting that, taking $X=R^{n}$, the notion of a
fuzzy convex sublattice also specializes to that of a ``classical'' convex
fuzzy set \cite[p.41]{Nguyen}. Fuzzy intervals and convex fuzzy sets in this
``classical'' sense have been studied extensively. It appears worthwhile to
study ``classical'' fuzzy intervals from the lattice theoretic point of view.
Conversely, they can serve as a source of inspiration for generalizations
(especially of convexity results) in the context of a general lattice
$(X,\sqsubseteq,\sqcup,\sqcap)$.

\end{document}